\documentclass[twocolumn,showpacs,aps,amsmath,amssymb,prc,superscriptaddress]{revtex4}

\usepackage{graphicx}% Include figure files
\usepackage{dcolumn}% Align table columns on decimal point
\usepackage{bm}% bold math
\usepackage{epstopdf} % needed for pdftex
\usepackage{color}

\newcommand{\Lam}{\ensuremath{\Lambda}}
\newcommand{\ALam}{\ensuremath{\bar{\Lambda}}}

\begin{document}

\title{Forward $\Lambda$ Production and Nuclear Stopping Power in $d$ + Au Collisions
at \mbox{$\sqrt{s_{NN}}$ = 200 GeV}}

\affiliation{Argonne National Laboratory, Argonne, Illinois 60439}
\affiliation{University of Birmingham, Birmingham, United Kingdom}
\affiliation{Brookhaven National Laboratory, Upton, New York 11973}
\affiliation{California Institute of Technology, Pasadena, California 91125}
\affiliation{University of California, Berkeley, California 94720}
\affiliation{University of California, Davis, California 95616}
\affiliation{University of California, Los Angeles, California 90095}
\affiliation{Carnegie Mellon University, Pittsburgh, Pennsylvania 15213}
\affiliation{University of Illinois at Chicago, Chicago, Illinois 60607}
\affiliation{Creighton University, Omaha, Nebraska 68178}
\affiliation{Nuclear Physics Institute AS CR, 250 68 \v{R}e\v{z}/Prague, Czech Republic}
\affiliation{Laboratory for High Energy (JINR), Dubna, Russia}
\affiliation{Particle Physics Laboratory (JINR), Dubna, Russia}
\affiliation{University of Frankfurt, Frankfurt, Germany}
\affiliation{Institute of Physics, Bhubaneswar 751005, India}
\affiliation{Indian Institute of Technology, Mumbai, India}
\affiliation{Indiana University, Bloomington, Indiana 47408}
\affiliation{Institut de Recherches Subatomiques, Strasbourg, France}
\affiliation{University of Jammu, Jammu 180001, India}
\affiliation{Kent State University, Kent, Ohio 44242}
\affiliation{Institute of Modern Physics, Lanzhou, China}
\affiliation{Lawrence Berkeley National Laboratory, Berkeley, California 94720}
\affiliation{Massachusetts Institute of Technology, Cambridge, MA 02139-4307}
\affiliation{Max-Planck-Institut f\"ur Physik, Munich, Germany}
\affiliation{Michigan State University, East Lansing, Michigan 48824}
\affiliation{Moscow Engineering Physics Institute, Moscow Russia}
\affiliation{City College of New York, New York City, New York 10031}
\affiliation{NIKHEF and Utrecht University, Amsterdam, The Netherlands}
\affiliation{Ohio State University, Columbus, Ohio 43210}
\affiliation{Panjab University, Chandigarh 160014, India}
\affiliation{Pennsylvania State University, University Park, Pennsylvania 16802}
\affiliation{Institute of High Energy Physics, Protvino, Russia}
\affiliation{Purdue University, West Lafayette, Indiana 47907}
\affiliation{Pusan National University, Pusan, Republic of Korea}
\affiliation{University of Rajasthan, Jaipur 302004, India}
\affiliation{Rice University, Houston, Texas 77251}
\affiliation{Universidade de Sao Paulo, Sao Paulo, Brazil}
\affiliation{University of Science \& Technology of China, Hefei 230026, China}
\affiliation{Shanghai Institute of Applied Physics, Shanghai 201800, China}
\affiliation{SUBATECH, Nantes, France}
\affiliation{Texas A\&M University, College Station, Texas 77843}
\affiliation{University of Texas, Austin, Texas 78712}
\affiliation{Tsinghua University, Beijing 100084, China}
\affiliation{Valparaiso University, Valparaiso, Indiana 46383}
\affiliation{Variable Energy Cyclotron Centre, Kolkata 700064, India}
\affiliation{Warsaw University of Technology, Warsaw, Poland}
\affiliation{University of Washington, Seattle, Washington 98195}
\affiliation{Wayne State University, Detroit, Michigan 48201}
\affiliation{Institute of Particle Physics, CCNU (HZNU), Wuhan 430079, China}
\affiliation{Yale University, New Haven, Connecticut 06520}
\affiliation{University of Zagreb, Zagreb, HR-10002, Croatia}

\author{B.I.~Abelev}\affiliation{University of Illinois at Chicago, Chicago, Illinois 60607}
\author{M.M.~Aggarwal}\affiliation{Panjab University, Chandigarh 160014, India}
\author{Z.~Ahammed}\affiliation{Variable Energy Cyclotron Centre, Kolkata 700064, India}
\author{B.D.~Anderson}\affiliation{Kent State University, Kent, Ohio 44242}
\author{D.~Arkhipkin}\affiliation{Particle Physics Laboratory (JINR), Dubna, Russia}
\author{G.S.~Averichev}\affiliation{Laboratory for High Energy (JINR), Dubna, Russia}
\author{Y.~Bai}\affiliation{NIKHEF and Utrecht University, Amsterdam, The Netherlands}
\author{J.~Balewski}\affiliation{Indiana University, Bloomington, Indiana 47408}
\author{O.~Barannikova}\affiliation{University of Illinois at Chicago, Chicago, Illinois 60607}
\author{L.S.~Barnby}\affiliation{University of Birmingham, Birmingham, United Kingdom}
% ...

\collaboration{STAR Collaboration}\noaffiliation

\date{December 17, 2007}

\begin{abstract}
We report the measurement of \Lam\ and \ALam\ yields and inverse slope
parameters in $d$ + Au collisions at $\sqrt{s_{NN}}$ = 200 GeV at forward and backward rapidities
($y$ = $\pm$ 2.75), using data from the STAR forward time projection chambers. The contributions of different processes to
baryon transport and particle production are probed exploiting the inherent asymmetry of 
the $d$ + Au system. Comparisons to model calculations show that the 
baryon transport on the deuteron side is consistent with multiple collisions 
of the deuteron nucleons with gold participants. On the gold side  
HIJING based models without a hadronic rescattering phase do not describe the measured particle yields while models that include target remnants or hadronic rescattering do. The multichain model can provide a good description of the net baryon density in $d$ + Au collisions at energies currently available at the BNL Relativistic Heavy Ion Collider, and the derived parameters of the model agree with those from nuclear collisions at lower energies.
\end{abstract}

\maketitle

\section{Introduction}

The production of strange baryons has been studied extensively in heavy-ion collisions
at the BNL Relativistic Heavy Ion Collider (RHIC) \cite{Adams:2005dq}. So far, these measurements
have been concentrated around midrapidity. In $d$ + Au collisions, the study of
particle production away from midrapidity is especially appealing since the inherent asymmetry
of these events allows the probing of different processes for particle production on the deuteron side (the side in the $d$ beam direction, positive rapidity) and on the gold side  (the side in the Au beam direction, negative rapidity) of the reaction. The deuteron side of the collision is expected to be  dominated
by multiple collisions of the incoming deuteron nucleons with gold participants and corresponding nuclear effects;  while on the gold side, final state rescattering and soft processes in the nuclear breakup might contribute significantly. By studying
the particle yields and the inverse slope parameters of \Lam\ and \ALam\ as a function of centrality, these effects are investigated. The centrality dependence of the yields can help illuminate the role of nuclear effects in the observed strangeness enhancement in Au + Au collisions \cite{Adler:2002uv, Adcox:2002au}. 

In addition to providing information about strange particle production in asymmetric collisions, 
\Lam\ and \ALam\ production at forward rapidities in $d$ + Au collisions also probes baryon 
transport and nuclear stopping power. The study of nuclear stopping power is a fundamental
issue in heavy-ion physics \cite{Busza:1983rj}, since this quantity is related to the amount of energy and baryon number that get 
transferred from the beam particles into the reaction zone. This influences the properties of possible new states of matter created in these collisions. For these studies, measurements at 
forward rapidities are crucial, since incomplete stopping is expected at collider energies \cite{Bjorken:1982qr}. This should lead to significant net baryon densities near beam rapidity; while at lower energies, the bulk of the net baryons are concentrated near midrapidity. At energies available at the CERN Super Proton Synchrotron (SPS), comprehensive studies of 
the rapidity distribution of net baryons in asymmetric collision systems demonstrate that the rapidity loss in these collisions depends mainly on the thickness of the nuclear target \cite{Alber:1997sn}. Recent theoretical work suggests that nuclear stopping may arise from gluon bremsstrahlung in cold nuclear matter \cite{Vitev:2007ve}.
A measurement of the mean rapidity loss of baryons in central Au + Au collisions at $\sqrt{s_{NN}}$ = 200 GeV \cite{Bearden:2004s} and the comparison to such measurements in central Pb + Pb collisions at the SPS \cite{Appelshauser:1998yb} indicate that a saturation of stopping is reached in central nucleus-nucleus reactions with respect to the center of mass energy at the top RHIC energy \cite{Bearden:2004s}. This is shown by the deviation from a linear scaling of the rapidity loss with collision energy observed at RHIC energies.

This paper presents the measurement of \Lam\ and \ALam\ particle yields and inverse slope parameters
at forward ($y$ = 2.75 $\pm$ 0.25)  and backward rapidity ($y$ = --2.75 $\pm$ 0.25)  in $d$ + Au collisions at $\sqrt{s_{NN}}$ = 200 GeV. By comparing the particle yields to  model calculations performed with AMPT \cite{Zhang:1999bd, Lin:2003ah}, EPOS \cite{Werner:2005jf}, HIJING \cite{Wang:1991ht} and HIJING/B$\bar{\text{B}}$ \cite{Vance:1999pr, Pop:2005uz}, information about the mechanisms for particle production in asymmetric collisions is gained. The net \Lam\ yield ($dN/dy$(\Lam ) -- $dN/dy$(\ALam)) as a function of centrality is compared to calculations based on the multichain model \cite{Date:1985sy} that was previously successfully applied to lower energy data. This comparison indicates that the baryon rapidity loss in $d$ + Au collisions at RHIC is consistent with the predictions of the multichain model. 

\section{Experimental Setup and Analysis Technique}

\begin{figure*}
	\includegraphics[width=\textwidth]{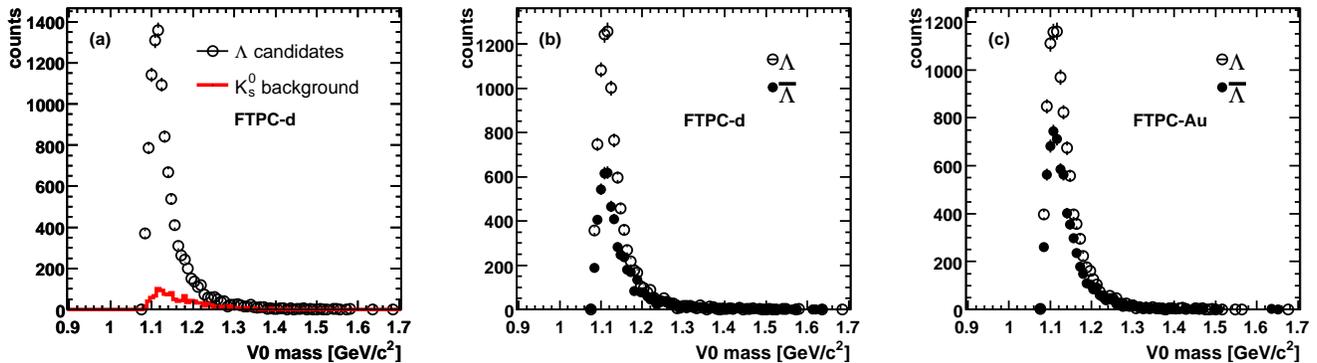}

	\caption{(Color online) a) Invariant mass distribution of \Lam\ candidates on the deuteron side 
	with estimated $K^0_s$ background distribution, b) and c) show the background subtracted
	\Lam\ and \ALam\ invariant mass distributions. The widths of the peaks are due 
	to the limited momentum resolution of the detectors, and are reproduced by simulations.}
	\label{fig:MinvPlots}
\end{figure*}

The data discussed here were taken with the solenoidal tracker (STAR) detector \cite{Ackermann:2002ad} at the RHIC accelerator
facility. The minimum bias trigger used for the data in this analysis required at least one beam momentum
neutron in the zero degree calorimeter (ZDC) in the Au beam direction. This trigger accepts 95$\pm$3\% of the
$d$ + Au hadronic cross section \cite{Adams:2003im}. The main detectors for the present analysis were the two radial-drift forward
time projection chambers (FTPCs) \cite{Ackermann:2002yx} that cover $2.5 < |\eta | < 4.0$ in pseudorapidity on
both sides of the interaction region. The detector that sits on the side of the interaction region the gold particles fly towards, and thus intercepts gold fragments, is referred to as FTPC-Au (negative rapidity). The 
detector on the deuteron side of the experiment is referred to as FTPC-d (positive rapidity). Data from the main TPC \cite{Anderson:2003ur} of STAR is used to determine the event vertex and
to provide a measure of the collision centrality, based on the charged particle multiplicity at midrapidity. 
This method of collision centrality determination avoids autocorrelations in the analysis, since different detectors are used for the measurements discussed here and for the
centrality definition. The minimum bias data set is thus divided into three centrality bins, as suggested in \cite{Kharzeev:2002ei}. The peripheral bin contains the 40\% -- 100\% most peripheral events, the mid-central bin contains 20\% -- 40\%, and the central bin contains the 20\% most central events, as defined by the charged particle multiplicity. The number of binary collisions, the number of $d$ and the number of Au participants for each of these bins are determined using Monte Carlo Glauber calculations incorporating the Hulth\'en wave function of the deuteron \cite{Miller:2007ri} . Table \ref{tab:Glauber} summarizes the Glauber calculation results. Contrary to the case of collisions of large, equal-sized nuclei, in $d$ + Au collisions the mean number of binary collisions is smaller than the mean number of participants since in most cases each Au nucleon only participates in one collision.  

\begin{table}
\begin{ruledtabular}
\begin{tabular}{lcc}
\rule{0mm}{3ex} centrality & \rule{3mm}{0mm}$\left< N_{part} \right>$ & $\left< N_{bin} \right>$ \\
\hline
\rule{0mm}{3ex} minimum bias& 8.3 $\pm$ 0.4 & 7.5 $\pm$ 0.4\\
\rule{0mm}{3ex} central (top 20\%) & 15.7$^{+1.2}_{-1.0}$& 15.0$^{+1.3}_{-0.9}$\\
\rule{0mm}{3ex} mid-central (20\% -- 40\%) & 11.2$^{+1.1}_{-1.0}$& 10.6$^{+0.8}_{-1.1}$\\
\rule{0mm}{3ex} peripheral (40\% -- 100\%) & 5.1 $\pm$ 0.4& 4.2 $\pm$ 0.4\\
\hline
\hline
\rule{0mm}{3ex} centrality & \rule{3mm}{0mm}$\left< N_{part, d} \right>$ & $\left< N_{part, Au} \right>$ \\
\hline
\rule{0mm}{3ex} minimum bias& 1.6 & 6.7\\
\rule{0mm}{3ex} central (top 20\%) & 2.0 & 13.7 \\
\rule{0mm}{3ex} mid-central (20\% -- 40\%) & 1.9 & 9.3\\
\rule{0mm}{3ex} peripheral (40\% -- 100\%) & 1.4 & 3.7\\
\end{tabular}
\end{ruledtabular}
\caption{Mean number of participants and mean number of binary collisions for minimum bias events and the three centrality classes, determined by Glauber calculations. Also given are the mean number of participants separated for the deuteron and the gold nucleus.}
\label{tab:Glauber}
\end{table}

After event selection cuts, which required a reconstructed primary event vertex along the beam axis within 50 cm of the center of the detector system, $10^7$ minimum bias events were accepted in the data sample. The vertex reconstruction efficiency 
was determined to be 93$\pm$1\% \cite{Adams:2003im}. Since the vertex reconstruction efficiency is a strong function of the number of tracks at midrapidity only peripheral events are affected by not reconstructed vertices. All particle spectra and yields are corrected for trigger and vertex finding efficiencies. 

The momentum resolution of the FTPCs was determined from simulations to be between about 10\% and 20\% for single charged tracks in the momentum and rapidity range covered by the present analysis. The momentum resolution for reconstructed \Lam\ and \ALam\ is on the order of 20\%. The binning in $p_T$ for the \Lam\ and \ALam\ spectra presented here is chosen accordingly.

In the FTPCs, \Lam\ and \ALam\ are reconstructed using their dominant decay modes $\Lam \rightarrow p\, \pi^-$
and $\ALam \rightarrow \bar{p}\, \pi^+$, which have a branching ratio of 64\%. \Lam\ candidates are identified 
via displaced vertices. The FTPCs measure a maximum of 10 points on a track. Due to the high momentum of particles in the forward rapidity region and consequently very similar energy loss $dE/dx$ of different particle species in the detector gas, particle identification via the specific energy loss is impossible for singly charged hadrons. 
Thus, \Lam\ candidates are formed from all pairs of one positive and one negative track which make up a possible decay vertex that is well separated from the main event vertex. These \Lam\ candidates are conventionally referred to as V0 due to their topology and charge. In the present analysis, a minimum separation of 20 cm is required. This large minimum decay length leads to a reduction in the overall reconstruction efficiency, which is corrected for in the analysis. 

In the case of \Lam\, the positive track is 
assumed to be a $p$, while the negative track is assumed to be a $\pi^-$. For \ALam\, the positive track is
assumed to be a $\pi^+$, while the negative track is assumed to be a $\bar{p}$. 
Since the most abundantly produced particle species are pions this lack of particle identification
introduces a considerable combinatoric background to the measurement. Strict cuts on the geometry of the assumed daughter tracks and the resulting \Lam\ candidate efficiently reduce this
background. The cut with the highest discriminating power was on the distance of closest
approach ($dca$) of the decay daughters to the primary vertex, which should be relatively 
small for the $p$ candidate and large for the $\pi$ candidate since the heavier decay daughter typically carries most of the momentum of the original particle and thus points back to the primary vertex while the lighter daughter does not. This cut selects track pairs originating from a decay vertex well separated from the primary vertex and in addition reduces the background from $K^{0}_{s}\, \rightarrow \, \pi^+\pi^-$ by favoring asymmetric decays. Additional cuts with high discriminating power were on the $dca$ of the resulting \Lam\ candidate and on the distance of the daughter tracks to the reconstructed decay vertex.

The remaining combinatoric background is determined by rotating the positive tracks by 180$^\circ$ with respect to the negative tracks in an event and then forming \Lam\ candidates using the same cuts as on real data. With this method the original spatial correlation of tracks is destroyed, removing real \Lam\ and \ALam\ from the sample. The combinatoric background, due to random track pairs that fulfill all analysis cuts, remains, since this depends on the track multiplicity and $dca$ distributions which are preserved in the rotated sample. The subtraction of this background results in a 15\% to 20\% correction. A mechanism leading to the loss of \Lam\ and \ALam\ signals is a possible misidentification of the charge of one of the two decay daughters. The probability increases with the particle momentum, so it is much more likely for the proton candidate than for the pion candidate. This effect manifests itself in like-sign track pairs that fulfill the geometrical requirements for the \Lam\ reconstruction. The size of this effect is determined with these like-sign pairs and is corrected for in the analysis. It is comparable in size to the remaining combinatoric background.  

The major source of background remaining after these cuts and corrections is
from $K^{0}_{s}\, \rightarrow \, \pi^+\pi^-$, where one of the two daughter
pions is assumed to be a proton. For the current analysis a full 
GEANT detector simulation with a HIJING \cite{Wang:1991ht} generated $K^0_s$ distribution, where both the  
transverse momentum and the rapidity spectra of the particles were taken from the event generator, was used to model this background. On the deuteron side it was verified that the $K^{0}_{s}$ yield taken from HIJING is in good agreement with $d$ + Au data in the kinematic region covered by the FTPCs. This was done using the same analysis procedure as for \Lam\ and \ALam, however using different cuts to allow for the different kinematics of the $K^{0}_{s}$ decay. 

For all V0s that pass the cuts, parameters such as the rapidity $y$, transverse momentum $p_T$, and invariant mass are calculated. In the present analysis, a rapidity range of $2.5 < |y| < 3.0$ was chosen since this range is fully within the FTPC acceptance
over the transverse momentum range of $0.5 < p_T < 2.0$ GeV/c used in the analysis.

Figure \ref{fig:MinvPlots}a) shows the invariant mass distribution for \Lam\ candidates on the deuteron 
side in the $p_T$ range from 0.5 GeV/c to 2.0 GeV/c for $d$ + Au minimum bias events. Also shown is the background contribution due to $K^{0}_{s}$ decays estimated from HIJING events. This background is subtracted, resulting in the
\Lam\ and \ALam\ invariant mass distributions shown for the deuteron side in Figure \ref{fig:MinvPlots}b) and for the
gold side in Figure \ref{fig:MinvPlots}c). On the gold side, the $p_T$ range is restricted to 0.7 GeV/c to 2.0 GeV/c, as discussed later in Section \ref{sec:Spectra}. From gaussian fits to the central part of the invariant mass distributions a mass of 1.116 GeV/c$^2$ was determined for both \Lam\ and \ALam\ on both sides of the collision, in good agreement with the literature value. The width of the mass peak, given by the $\sigma$ of the fit, is 24 MeV/c$^2$ for the deuteron side and 26 MeV/c$^2$ for the gold side, driven entirely by the detector resolution. The reconstructed mass is independent of centrality, but shows a slight $p_T$ dependence due to the $p_T$ dependent detector resolution. The variation over the studied transverse momentum range is around 10 MeV/c$^2$, with lower values at low $p_T$ and higher values at high $p_T$. The observed invariant mass distributions are reproduced by a full GEANT simulation taking into account the detector response. 

The raw particle yields are extracted by summing up the bin contents of the background-subtracted invariant mass distributions from 1.08 GeV/c$^2$ to 1.24 GeV/c$^2$. To get from the raw particle yields to corrected yields, the acceptance and the efficiency for \Lam\ and \ALam\ has to be determined. This is done by embedding into real $d$ + Au events simulated \Lam\ decays that were run through a GEANT model of the detector and a simulator of the FTPC response. The reconstruction efficiency for \Lam\ and \ALam\ in the range $2.5 < |y| < 3.0$ and 0.5 GeV/c $< p_T <$ 2.0 GeV/c is $\sim$ 6\% with a small dependence on $p_T$. This number includes the effect of detector acceptance
and the analysis cuts used. It is dominated by the requirement of a well-separated decay vertex. The branching ratio of the decay into charged particles is 64\%, leading to an overall efficiency of $\sim$ 4\%.

\section{Particle Spectra and Yields}
\label{sec:Spectra}

\begin{figure}
   \includegraphics[width=0.5\textwidth]{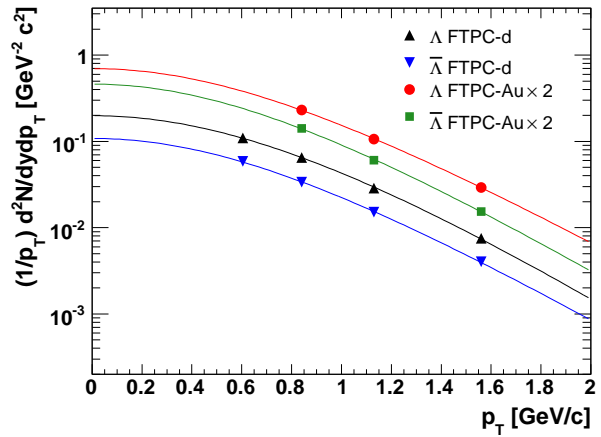} 
   \caption{(Color online) \Lam\ and \ALam\ spectra on the deuteron and on the gold side in $d$ + Au minimum bias collisions. The data points on the gold side are multiplied by 2 for better visibility. The statistical errors are smaller than the points marking the measurements. The curves show a fit with a Boltzmann function in transverse mass to the data points.}
   \label{fig:Spectra}
\end{figure}

Transverse momentum spectra for \Lam\ and \ALam\ in $d$ + Au minimum bias collisions at $\sqrt{s_{NN}}$ = 200 GeV are shown in Figure \ref{fig:Spectra} for both sides of the collision. An incorrect treatment of defective electronics in FTPC-Au during data production led to the inclusion of noisy electronics channels in the data analysis, affecting the measurements at low $p_T$ in particular.  This is due to an excess of low $p_T$ tracks that fulfill the $dca$ cuts for decay daughters, which manifests itself in a shifting of the reconstructed invariant mass at low $p_T$. Thus the region below $p_T$ = 0.7 GeV/c is excluded from the analysis on the Au side. Also shown are fits to the data with a Boltzmann distribution in transverse mass $m_T$,
\begin{equation}
\frac{1}{2\pi p_T} \frac{d^2N}{dydp_T} = C\, m_T\, exp(-m_T/T)
\end{equation}
where $T$ is the inverse slope parameter of the spectrum and $C$ is the overall normalization.
The spectra on both collision sides agree well with the assumed exponential behavior.
\begin{figure}
   \includegraphics[width=0.5\textwidth]{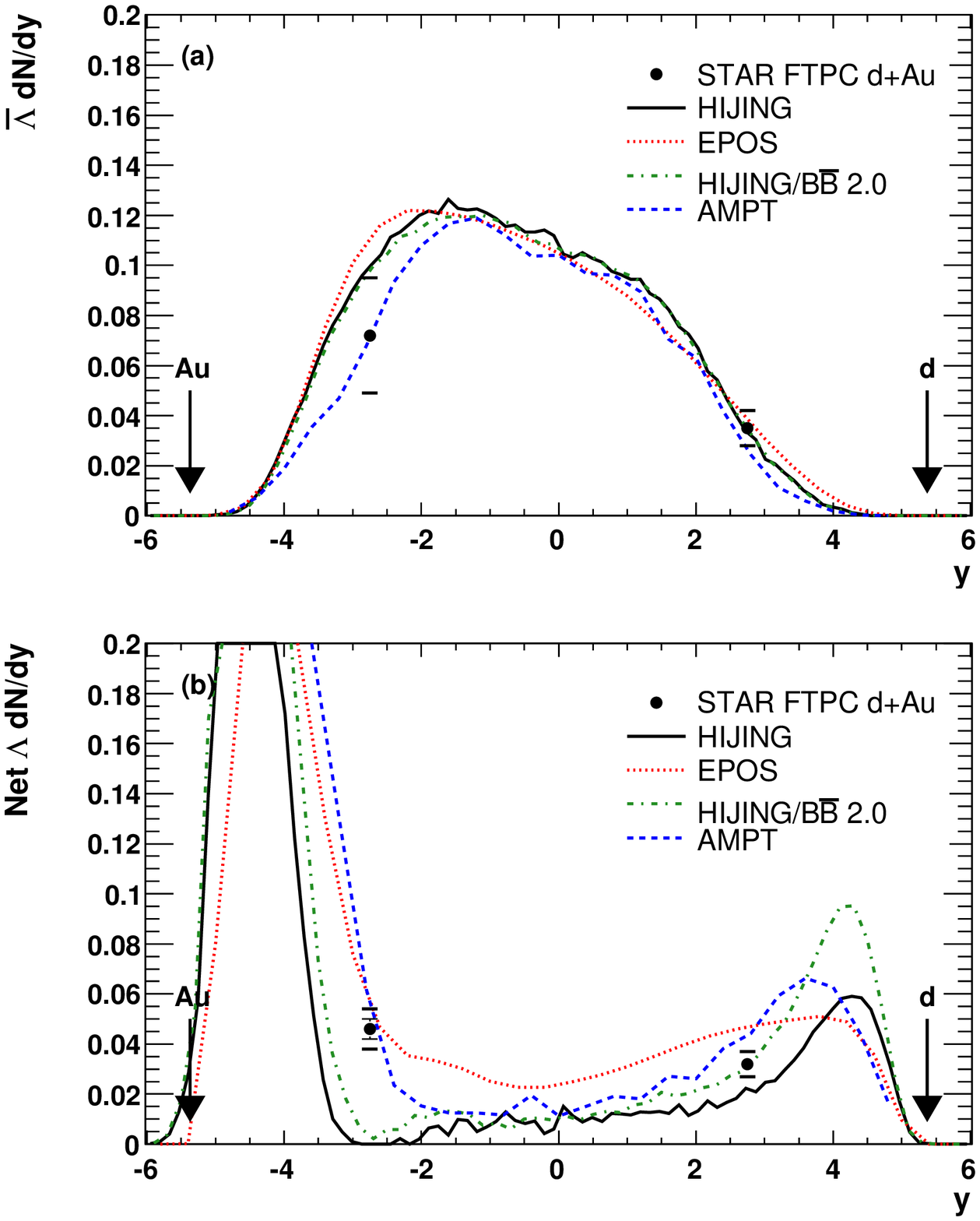} 
   \caption{(Color online) (a) Comparison of the measured \ALam\ yield with model calculations. \\
   (b) Comparison of the net \Lam\ yield with model calculations.\\
   Statistical errors are shown as vertical error bars, the vertical caps show the quadratic sum of statistical and systematic errors including the overall normalization uncertainty. In both panels the target and projectile beam rapidities are indicated by arrows.}
   \label{fig:ModelCompare}
\end{figure}

From the fits the total particle yield in the rapidity range $2.5 < |y| < 3.0$ is extrapolated and the inverse slope parameters are extracted. The missing low $p_T$ measurement in the $p_T$ spectra on the Au side leads to an additional systematic error both in the yield and the inverse slope parameter. The \Lam\ and \ALam\ yields as well as the inverse slope parameters are determined for minimum bias events and the three individual centrality classes: central (0\% -- 20\%), mid-central (20\% -- 40\%) and peripheral (40\% -- 100\%) events. The particle yields are corrected for acceptance, efficiency and feed-down from weak decays of hyperons with higher mass. The feed-down contribution is taken from HIJING simulations. The fraction of detected \Lam\ and \ALam\ particles originating from decays of higher mass hyperons was determined to be 0.1 $\pm$ 0.03. This number includes the differences in reconstruction efficiencies for primary \Lam\ (\ALam) and for \Lam\ (\ALam) from hyperon decays due to their displaced production point. As usual in heavy-ion collisions no correction is applied for the contribution from $\Sigma^0$ decays. Thus all quoted \Lam\ yields consist of the contributions of primary \Lam\ and $\Sigma^0$. Table \ref{tab:dNdy} summarizes the particle yields, while Table \ref{tab:InvSlope} shows the inverse slope parameters determined from the Boltzmann fits to the spectra as well as the $\ALam/\Lam$ yield ratio determined from the particle yields. Within the assumption of a Boltzmann distribution the observed slope parameters translate into mean transverse momenta $\left<p_T\right>$ between 0.74 GeV/c and 0.82 GeV/c. Within that assumption the fraction of the total yield that is covered by the measurement is $\sim 64\%$ on the $d$ side and $\sim 43\%$ on the Au side.

%\squeezetable

\begin{table*}
\begin{ruledtabular}
\begin{tabular}{l|c|c|c}
\rule{0mm}{3ex} centrality & \Lam\ $dN/dy$ & \ALam\ $dN/dy$ & net \Lam\ $dN/dy$ \\
\hline
\multicolumn{4}{c}{\rule{0mm}{2.7ex} deuteron side ($y$ = 2.75)}\\
\hline
\rule{0mm}{2.7ex} min. bias & $0.067 \pm 0.001\, (\text{stat})\, ^{+0.010}_{-0.009}\, \text{(syst)}$ & $0.035 \pm 0.001\, (\text{stat})\, ^{+0.006}_{-0.005}\, \text{(syst)}$ & $0.032\pm 0.002\, (\text{stat})\, \pm 0.004\, \text{(syst)}$ \\
\rule{0mm}{2.7ex} top 20\% &$0.106 \pm 0.003\, (\text{stat})\, ^{+0.016}_{-0.014}\, \text{(syst)}$ &$0.054 \pm 0.002\, (\text{stat})\, ^{+0.010}_{-0.008}\, \text{(syst)}$ &$0.052\pm 0.004\, (\text{stat})\, \pm 0.007\, \text{(syst)}$ \\
\rule{0mm}{2.7ex} 20\% -- 40\% &$0.094 \pm 0.003\, (\text{stat})\, ^{+0.014}_{-0.013}\, \text{(syst)}$&$0.047 \pm 0.002\, (\text{stat})\, ^{+0.009}_{-0.007}\, \text{(syst)}$&$0.047\pm 0.004\, (\text{stat})\, \pm 0.006\, \text{(syst)}$\\
\rule{0mm}{2.7ex} 40\% -- 100\%&$0.045 \pm 0.001\, (\text{stat})\, ^{+0.007}_{-0.006}\, \text{(syst)}$&$0.025 \pm 0.001\, (\text{stat})\, \pm 0.004\, \text{(syst)}$&$0.020\pm 0.002\, (\text{stat})\, \pm 0.003\, \text{(syst)}$\\
\hline
\multicolumn{4}{c}{\rule{0mm}{2.7ex} gold side ($y$ = -2.75)}\\
\hline
\rule{0mm}{2.7ex} min. bias & $0.118 \pm 0.004\, (\text{stat})\, ^{+0.030}_{-0.028}\, \text{(syst)}$&$0.072 \pm 0.002\, (\text{stat})\, \pm 0.022\, \text{(syst)}$ &$0.046\pm 0.004\, (\text{stat})\, \pm 0.006\, \text{(syst)}$\\
\rule{0mm}{2.7ex} top 20\% &$0.294 \pm 0.017\, (\text{stat})\, ^{+0.074}_{-0.070}\, \text{(syst)}$ &$0.176 \pm 0.010\, (\text{stat})\, \pm 0.054\, \text{(syst)}$ &$0.118\pm 0.020\, (\text{stat})\, \pm 0.015\, \text{(syst)}$ \\
\rule{0mm}{2.7ex} 20\% -- 40\% &$0.163 \pm 0.008\, (\text{stat})\, ^{+0.041}_{-0.039}\, \text{(syst)}$&$0.096 \pm 0.005\, (\text{stat})\, \pm 0.029\, \text{(syst)}$&$0.067\pm 0.009\, (\text{stat})\, \pm 0.009\, \text{(syst)}$\\
\rule{0mm}{2.7ex} 40\% -- 100\%&$0.048 \pm 0.002\, (\text{stat})\, \pm 0.012\, \text{(syst)}$&$0.031 \pm 0.002\, (\text{stat})\, \pm 0.009\, \text{(syst)}$ &$0.017\pm 0.003\, (\text{stat})\, \pm 0.003\, \text{(syst)}$\\

\end{tabular}
\end{ruledtabular}
\caption{Corrected yields of \Lam, \ALam\ and net \Lam\ on both sides of the collision. In addition to the quoted systematic errors there is an overall normalization uncertainty of 10\% on the particle yields.}
\label{tab:dNdy}
\end{table*}

\begin{table*}
\begin{ruledtabular}
\begin{tabular}{l|c|c|c}
\rule{0mm}{3ex} centrality & \Lam\ inverse slope [GeV] & \ALam\ inverse slope [GeV]& $\ALam/\Lam$ yield ratio\\
\hline
\multicolumn{4}{c}{\rule{0mm}{2.7ex}deuteron side ($y$ = 2.75)}\\
\hline
\rule{0mm}{2.7ex} min. bias &  $0.209 \pm 0.003\, (\text{stat})\, \pm 0.009\, \text{(syst)}$ & $0.210 \pm 0.004\, (\text{stat})\, \pm 0.009\, \text{(syst)}$&$0.52 \pm 0.02\, (\text{stat}) \pm 0.04\, (\text{syst})$\\ 
\rule{0mm}{2.7ex} top 20\% &$0.221 \pm 0.005\, (\text{stat})\, \pm 0.010\, \text{(syst)}$ &$0.224 \pm 0.007\, (\text{stat})\, \pm 0.010\, \text{(syst)}$&$0.51 \pm 0.02 \, (\text{stat})\, \pm  0.05\, (\text{syst})$\\
\rule{0mm}{2.7ex} 20\% -- 40\%  &$0.208 \pm 0.005\, (\text{stat})\, \pm 0.010\, \text{(syst)}$ &$0.213 \pm 0.007\, (\text{stat})\, \pm 0.010\, \text{(syst)}$&$0.50 \pm 0.03 \, (\text{stat})\, \pm  0.05\, (\text{syst})$\\
\rule{0mm}{2.7ex} 40\% -- 100\% &$0.202 \pm 0.004\, (\text{stat})\, \pm 0.009\, \text{(syst)}$ & $0.199 \pm 0.005\, (\text{stat})\, \pm 0.009\, \text{(syst)}$&$0.56 \pm 0.03 \, (\text{stat})\, \pm  0.05\, (\text{syst})$\\
\hline
\multicolumn{4}{c}{\rule{0mm}{2.7ex}gold side ($y$ = -2.75)}\\
\hline
\rule{0mm}{2.7ex} min. bias  &$0.219 \pm 0.005\, (\text{stat})\, \pm 0.013\, \text{(syst)}$ &$0.206 \pm 0.005\, (\text{stat})\, \pm 0.012\, \text{(syst)}$&$0.61 \pm 0.03\, (\text{stat}) \pm 0.05\, (\text{syst})$\\
\rule{0mm}{2.7ex} top 20\% &$0.217 \pm 0.006\, (\text{stat})\, \pm 0.013\, \text{(syst)}$ &$0.210 \pm 0.007\, (\text{stat})\, \pm 0.012\, \text{(syst)}$&$0.60 \pm 0.05 \, (\text{stat})\, \pm  0.05\, (\text{syst})$\\
\rule{0mm}{2.7ex} 20\% -- 40\%  &$0.218 \pm 0.007\, (\text{stat})\, \pm 0.013\, \text{(syst)}$ &$0.204 \pm 0.008\, (\text{stat})\, \pm 0.012\, \text{(syst)}$&$0.59 \pm 0.04 \, (\text{stat})\, \pm  0.05\, (\text{syst})$ \\
\rule{0mm}{2.7ex} 40\% -- 100\% &$0.221 \pm 0.007\, (\text{stat})\, \pm 0.013\, \text{(syst)}$ &$0.201 \pm 0.008\, (\text{stat})\, \pm 0.011\, \text{(syst)}$&$0.65 \pm 0.05 \, (\text{stat})\, \pm  0.05\, (\text{syst})$  \\

\end{tabular}
\end{ruledtabular}
\caption{Inverse slope parameters determined from Boltzmann fit in $m_T$ for \Lam\ and \ALam\ on both sides of the collision and $\ALam/\Lam$ yield ratios.}
\label{tab:InvSlope}
\end{table*}

The systematic errors quoted for the results include several contributions. These contributions are the cut parameters, the efficiency determination from embedding, background normalization and feed down corrections. The size of each contribution is obtained from the size of effects on the results from variations of cuts and normalizations and from a comparison of measurements in different sub-regions of the detectors. The dominating contributions are from uncertainties introduced by the cut selection and from the efficiency determination. This contribution is up to 12\% on the deuteron side and as large as 20\% on the Au side. Since both the background and feed down contributions in the raw signal are relatively small, the large uncertainties on their normalizations do not lead to sizeable systematics on the extracted yields. Systematics are evaluated separately for the particle yields and the $\ALam /\Lam$ ratio. In the determination of the net \Lam\ yield and of the $\ALam /\Lam$ ratio a significant fraction of the systematic effects cancel, leading to smaller overall systematics in these quantities compared to the \Lam\ and \ALam\ yields. Especially in the case of the yields the systematics on the Au side are considerably larger than on the $d$ side due to the aforementioned problems with the treatment of noisy electronics channels. 

As an additional systematic check the fits to the spectra were also performed with an exponential function in $m_T$ of the form
\begin{equation}
\frac{1}{2\pi p_T} \frac{d^2N}{dydp_T} = C'\, exp(-m_T/T')
\end{equation}
where $T'$ is the inverse slope parameter of the exponential function and $C'$ is the overall normalization. The yields extracted with this exponential function are consistent with the yields extracted based on the assumption of a Boltzmann distribution. Our systematic errors do not include yield variations due to different spectrum functions.

In addition to the systematic errors associated with the analysis there is a 10\% overall normalization uncertainty in the particle yields \cite{Adams:2003im}. This uncertainty is included in the systematic errors shown in the model comparison figures and in the baryon transport study.

To investigate the mechanisms that drive particle production at forward and backward rapidity in $d$ + Au collisions, the measured minimum bias yields are compared to a variety of model calculations. Figure \ref{fig:ModelCompare} shows the measured \ALam\ and net \Lam\ yield compared to model calculations. While the \ALam\ yield is sensitive to the amount of \Lam -\ALam\ pair production, the net \Lam\ yield is strongly influenced by baryon number transport in the collision. Incoming nucleons can be transformed to \Lam\ via the associated production of strangeness, $N + N \rightarrow N + \Lam + K$, leading to a strong correlation of net \Lam\ and net baryon numbers. 

The four models used are based on different principles. HIJING \cite{Wang:1991ht} treats nucleus-nucleus collisions as a superposition of individual nucleon-nucleon collisions with a parametrization of nuclear effects such as shadowing and does not include final state effects such as scattering in the hadronic phase. The HIJING/B$\bar{\text{B}}$ model \cite{Vance:1999pr, Pop:2005uz} is based on HIJING but includes baryon junction interactions for increased baryon number transport. These gluonic structures facilitate baryon number transfer over large rapidity intervals \cite{Kharzeev:1996sq}. AMPT \cite{Zhang:1999bd, Lin:2003ah} is a multi-phase model that includes a HIJING-like treatment of initial nucleon-nucleon reactions as well as a later rescattering phase. EPOS \cite{Werner:2005jf} is a phenomenological approach based on a parton model. It  incorporates nuclear effects via parton ladders and target and projectile remnants. None of the model calculations have been tuned to provide agreement with the data presented here, except in the case of HIJING/B$\bar{\text{B}}$. Here the version with a string tension constant of $\kappa_i$ = 1.5 GeV/fm is used since this showed the best agreement with the \ALam\ yields on both the $d$ and the Au side and thus with the observed \Lam-\ALam\ pair production. For all model comparisons discussed below the systematic errors in the model calculations are not considered.

On the deuteron side, the HIJING description of several consecutive nucleon-nucleon reactions that take place as the nucleons of the deuteron pass through the gold nucleus is assumed to be appropriate. It is expected that all models should give good descriptions of particle production on the deuteron side. On the gold side, however, effects associated with final state rescattering and the breakup of the target nucleus are expected to play a major role, so the AMPT and EPOS models are expected to show a better performance.

Comparing the \ALam\ measurements with the rapidity distributions predicted by the models, shown in Figure \ref{fig:ModelCompare}a), it is apparent that indeed the deuteron side is very well described, with all models yielding the same result. On the gold side, AMPT is below the other three models. It agrees best with the data, however, also the two HIJING models and EPOS are consistent with the measurement. In general, all models used give a fair description of the \ALam\ yield and thus of the \Lam -\ALam\ pair production in minimum bias $d$ + Au collisions.

Larger differences are seen for the net \Lam\ yield shown in Figure \ref{fig:ModelCompare}b), which is very sensitive to baryon transport, since the \Lam\ has two quarks in common with a nucleon and thus can easily be produced from a nucleon via associated production of strangeness. On the deuteron side,  HIJING/B$\bar{\text{B}}$ shows the best agreement with the data, suggesting multiple nucleon-nucleon collisions with additional baryon transport are an appropriate description of the deuteron side of the collision. On the gold side significant differences between the models are apparent. Neither HIJING nor HIJING/B$\bar{\text{B}}$ reproduce the measured net \Lam\ yield at negative rapidity, while AMPT and EPOS do. This suggests that target related effects, as implemented in AMPT and EPOS, have a strong influence on strangeness production on the Au side. It appears that at least either a final state rescattering phase, as implemented in AMPT, or the inclusion of target remnants in EPOS, is necessary to reproduce the observed net \Lam\ yield on the gold side of the reaction.

\begin{figure*}
\includegraphics[width=\textwidth]{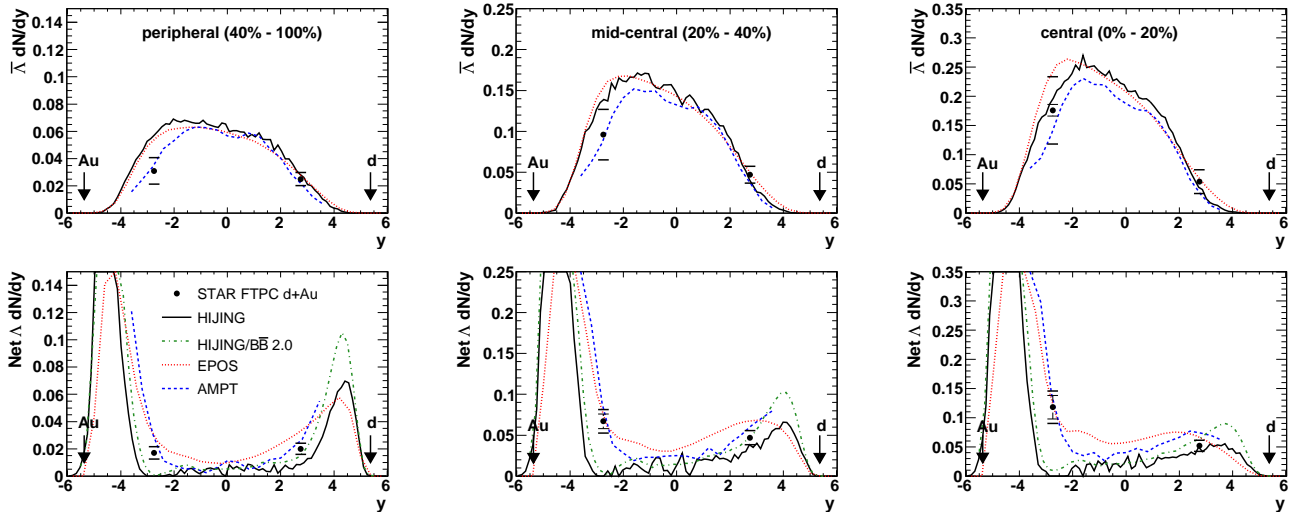}
\caption{(Color online) Comparison of \ALam\ and net \Lam\ yields to model calculations for all three centrality classes. Statistical errors are shown as vertical error bars, the vertical caps show the quadratic sum of statistical and systematic errors. Beam rapidity is indicated by arrows.}
\label{fig:CentralityEvolution}
\end{figure*}

Figure \ref{fig:CentralityEvolution} shows the \ALam\ and net \Lam\ yields for the three separate centrality bins compared to calculations with HIJING, HIJING/B$\bar{\text{B}}$ (net \Lam\ only), AMPT and EPOS. The agreement of the models with the \ALam\ yields on the deuteron side observed for the minimum bias dataset holds for centrality selected collisions. The evolution of the \ALam\ yield as a function of centrality on the Au side exceeds the increase predicted by the HIJING model. While the \ALam\ yield in peripheral events agrees best with the AMPT prediction, the central result is consistent with all three models. In general, the yield increase on the gold side significantly exceeds the yield increase on the deuteron side with increasing collision centrality. The behavior of the net \Lam\ yield as a function of centrality is consistent with the observations in minimum bias collisions. While HIJING/B$\bar{\text{B}}$ provides the best match to the data on the deuteron side, the gold side is not described by the HIJING models. EPOS and AMPT are able to describe the centrality evolution of the net \Lam\ yield on the Au side. On the deuteron side, all models indicate a transition from large transparency to significant stopping in the probed centrality range. This behavior will be further investigated in Section \ref{sec:Stopping}.

\begin{figure}
   \includegraphics[width=0.5\textwidth]{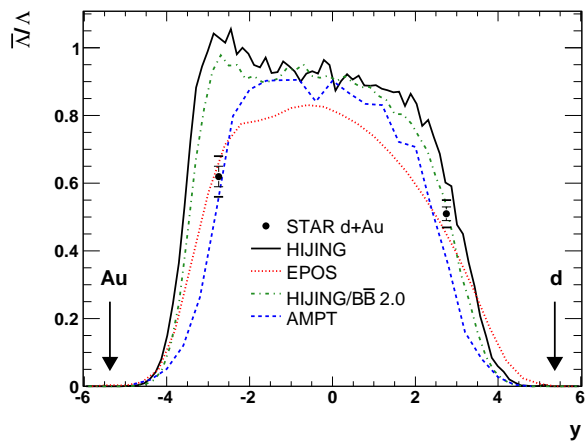} 
   \caption{(Color online) Minimum bias $\ALam/\Lam$ ratio compared to model calculations. On the deuteron side  HIJING/B$\bar{\text{B}}$ shows the best agreement with the results, while on the Au side only AMPT and EPOS give a satisfactory description of the data.}
   \label{fig:ModelRatio}
\end{figure}

\begin{figure*}
   \includegraphics[width=0.9\textwidth]{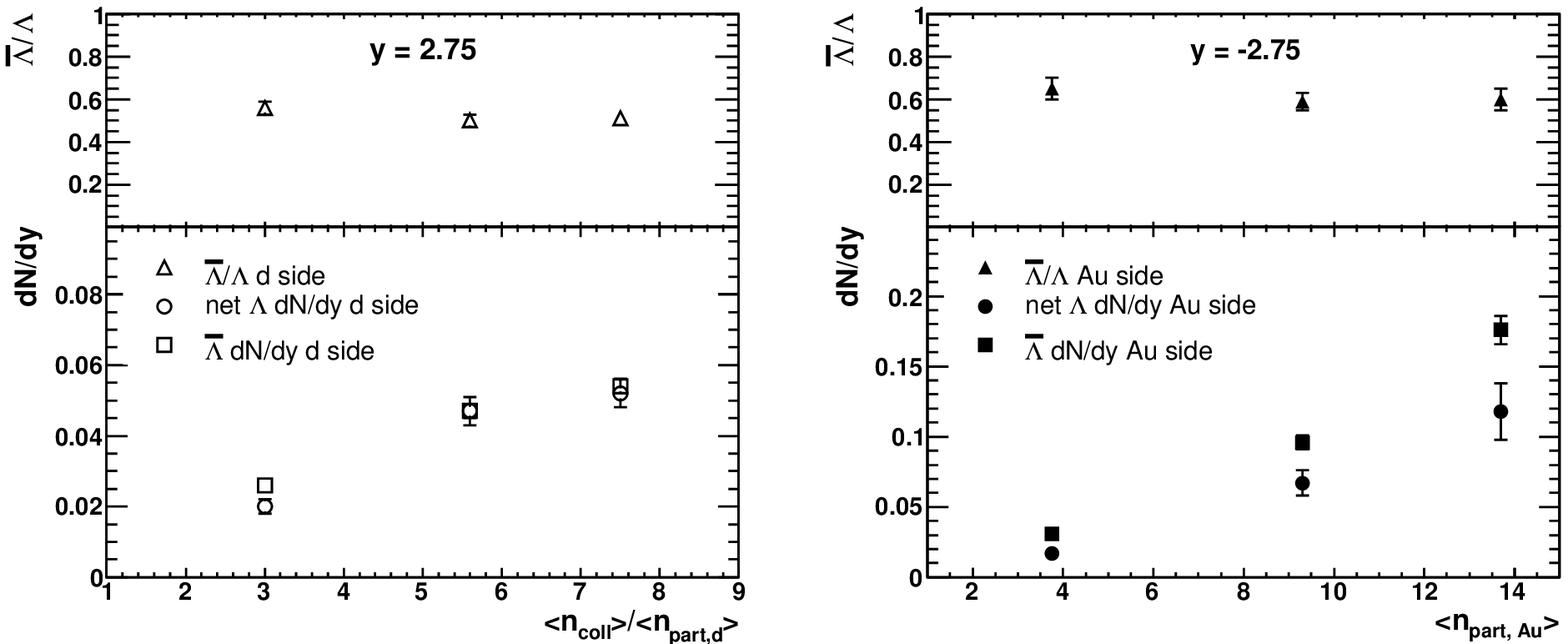} 
   \caption{\ALam/\Lam\ ratio and net \Lam\ and \ALam\ yields as a function of collision centrality on both the deuteron (left) and the gold side (right). On the deuteron side, centrality is expressed by the number of collisions per deuteron participant, while on the gold side the number of Au participants is chosen.  Only statistical errors are shown. The increase in baryon number transport with centrality, shown by the net \Lam\ yield, is matched by the increase of \ALam -\Lam\ pair production, thus keeping the \ALam/\Lam\ ratio constant over a wide centrality range.}
   \label{fig:RatioYield}
\end{figure*}

The minimum bias $\ALam/\Lam$ yield ratio together with predictions from the four models discussed above is shown in Figure \ref{fig:ModelRatio}. As for the net \Lam\ yields, all models are close to the data on the deuteron side with  HIJING/B$\bar{\text{B}}$ showing the best match. On the gold side AMPT and EPOS, which both incorporate nuclear effects, can reproduce the measurement.

An interesting feature of the centrality dependence of the $\ALam/\Lam$ ratio, shown in the upper panels of Figure \ref{fig:RatioYield}, is that while the measured net \Lam\ yields change significantly with centrality on both sides of the collision in the measured rapidity bins, the ratio stays constant within statistical errors. This shows that the increase in baryon stopping with collision centrality is not reflected in a decrease of the anti-baryon to baryon ratio, at least not in the hyperon sector. While the net \Lam\ yield, given by the difference of \Lam\ and \ALam\ yield, is directly linked to the amount of baryon number transport in the reaction, the anti-particle to particle ratio is influenced by baryon transport and \ALam-\Lam\ pair production. The centrality independence of the ratio suggests that baryon number transport and pair production increase in a similar way with increasing collision centrality and thus with the amount of nuclear material traversed by the projectile. This is shown in the lower panel of Figure \ref{fig:RatioYield} with the net \Lam\ and the \ALam\ yield as a function of centrality on both collision sides. This is in line with previous $p$ + $A$ measurements with a proton beam of up to 300 GeV on a fixed target which showed very similar \ALam/\Lam\ ratios for $p$ + Be and $p$ + Pb reactions \cite{Skubic:1978fi}. These measurements were performed on the projectile ($p$) side of the collision as a function of $x_F$, defined as $p_{||,\Lambda} / p_{max}$, where $p_{||,\Lambda}$ is the longitudinal component of the \Lam\ momentum and $p_{max}$ is the maximal possible longitudinal momentum of the \Lam\ in the center of mass frame (of a nucleon-nucleon system). The $x_F$ range of these measurements was $\sim$ 0.2 to $\sim$ 0.4, compared to an $x_F$ of $\sim$ 0.1 for the $d$ side data presented here. 

From the inverse slope parameters listed in Table \ref{tab:InvSlope} it can be seen that the inverse slopes of both \Lam\ and \ALam\ show a collision side dependent behavior with centrality. Within statistical errors, the inverse slope parameter does not change with the number of collisions on the gold side. On the deuteron side, an increase with centrality and thus with the number of nucleon-nucleon collisions the deuteron constituents participate in is observed. This effect is attributed to an increase of the mean transverse momentum of the particles in each of the subsequent collisions the deuteron participants suffer on their way through the gold nucleus, and agrees with the picture of the deuteron side of the reaction discussed above. This observation is in contrast to the behavior of inclusive charged hadrons in $d$ + Au collisions where an increase of the mean $p_T$ with centrality was observed on the Au side, while no centrality dependence was found on the deuteron side \cite{Joern}. This suggests a difference in the behavior of hyperons and charged hadrons, mainly pions, in the dynamical evolution of the nuclear collision.

From the \Lam\ and \ALam\ spectra and yields at forward and backward rapidities in $d$ + Au collisions, it is indicated that the deuteron side of the collision is dominated by multiple consecutive nucleon-nucleon collisions of participants of the incoming deuteron with gold nucleons. On the gold side, the HIJING models can not reproduce the observed net \Lam\ production, while models including nuclear effects can. This situation is different from that found for inclusive charged hadron yields in $d$ + Au collisions at mid-and forward rapidity, where no significant sensitivity to various model calculations has been observed \cite{Joern}.
Studies at midrapidity suggest that more differential measurements, like the ratio of particle production at backward rapidity to forward rapidity as a function of $p_T$, can provide additional information on the relative contributions of various physical processes to particle production \cite{Abelev:2006pp,Adams:2004dv}. Although such a study is beyond the scope of the present work, there is a consistent picture emerging regarding the model preference of the $d$ + Au data at both mid- and forward rapidity. Specifically, midrapidity studies do not support models based on incoherent initial multiple partonic scattering and independent fragmentation, such as HIJING. The EPOS model, which provides a good match to the measurements on \Lam\ production presented here was also found to explain the data at midrapidity in $d$ + Au collisions across many observables \cite{Abelev:2006pp, Adams:2006nd}.

\section{Baryon Transport and Nuclear Stopping Power}
\label{sec:Stopping}

The discussions in the previous section can be extended to a study of baryon transport in $d$ + Au collisions based on comparisons to the Multi-Chain Model (MCM) \cite{Date:1985sy}. This model predicts the baryon number transport in nuclear collisions based on simple assumptions. To do this, the net \Lam\ yields presented here have to be related to the total number of net baryons in the corresponding rapidity ranges. This is done using model calculations performed with HIJING/B$\bar{\text{B}}$ \cite{Topor-Pop:priv}. On the Au side of the collision there are clearly some issues with the description of the net \Lam\ yield by the HIJING/B$\bar{\text{B}}$ model, as discussed above. Thus only the deuteron side in the rapidity range from 0 to 4.0 is used to extract the following correspondence: net baryons = (10 $\pm$ 1) $\times$ net \Lam.  For the purpose of this discussion, this is assumed to be valid also on the Au side of the reaction. Since the MCM predicts net baryon yields, the model output is scaled by this parameter before comparing to the data presented here. The model curves are obtained by coupling the MCM as described in \cite{Date:1985sy} with probability distributions for the number of binary $N+N$ collisions obtained from Glauber calculations. 

In the literature nuclear stopping power is usually described by the mean rapidity loss of incoming baryons in the nuclear collisions \cite{Busza:1983rj}, 
\begin{equation}
\delta y = y_{beam} - \left< y \right>,
\end{equation}
where $y_{beam}$ is the beam rapidity and $\left< y \right>$ is the mean rapidity of the projectile baryons after the collision. In the MCM, the mean rapidity loss is related to the single phenomenological parameter $\alpha$ by 
\begin{equation}
\delta y = (n_{coll} - 1)/\alpha + 1,
\end{equation}
where $n_{coll}$ is the number of collisions the incoming nucleon suffers. The distribution of the number of collisions and the probabilities for one or two nucleons of the projectile participating in the reaction are determined with Glauber calculations using the multiplicity based centrality definitions used in the data analysis. The parameter $\alpha$ was originally extracted from an analysis of results on $p + A \rightarrow p + X$ at 100 GeV fixed-target beam energy \cite{Barton:1982dg}, with a result of $\alpha = 3 \pm 1$.    

\begin{figure}
   \includegraphics[width=0.5\textwidth]{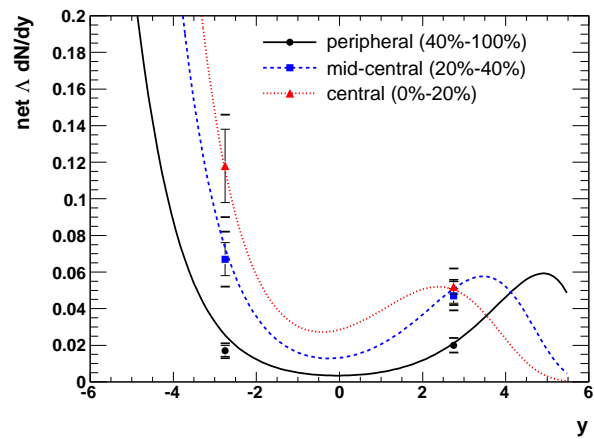} 
   \caption{(Color online) Net \Lam\ $dN/dy$ for central, mid-central and peripheral events on both the deuteron and the Au side of the collision. The data are compared to calculations of the distribution of net baryons obtained with the Multichain model \cite{Date:1985sy} with $\alpha$ = 2.9, scaled by 0.1 to account for the conversion from net baryons to net \Lam. An overall scale uncertainty of 10\% on the model curves from this conversion is not shown. See text for details.}
   \label{fig:TransportCentrality}
\end{figure}

Figure \ref{fig:TransportCentrality} shows the measured net \Lam\ yields on both sides of the collision for all three centrality bins together with predictions based on the MCM using $\alpha = 2.9$. Uncertainties of the overall scale of the model curves due to the conversion from net baryons to net \Lam\ are on the order of 10\% and are not shown here. The value of $\alpha = 2.9$ adopted for the figure is the best fit to the results. Good fits are provided in the range of $\alpha = 2.9 \pm 0.5$, ignoring uncertainties stemming from the conversion from net baryons to net \Lam. The data show good agreement with the MCM independent of collision centrality. The range for the model parameter $\alpha$ supported by the data is driven mostly by the measurements on the deuteron side. On the gold side the net baryon yield is dominated by baryons transported from the target rapidity. The rapidity distribution of baryons on the Au side is only weakly dependent on $\alpha$, since most participating target (gold) nucleons only suffer one single collision in the reaction and thus only baryons transported from the projectile side to the target side contribute to an $\alpha$ dependence. The model parameter extracted from the net \Lam\ data in $d$ + Au collisions at $\sqrt{s_{NN}}$ = 200 GeV is consistent with that obtained from $p$ + $A$ collisions at 100 GeV fixed target energy. 

The good agreement of the MCM with a common parameter for reactions with more than an order of magnitude different center of mass energy suggests that the rapidity loss of the incoming baryons in $p$($d$) + A collisions and thus the nuclear stopping power is largely independent of beam energy and to a good approximation only a function of the number of collisions over a wide energy range.

In central Au + Au collisions at $\sqrt{s_{NN}}$ = 200 GeV a saturation of the stopping power with energy has been observed that was not seen in previous measurements at lower energy \cite{Bearden:2004s}. From the expectations of MCM with $\alpha = 2.9$ it appears that the rapidity loss in central nucleus-nucleus collisions is lower than that in $d$ + Au collisions for a comparable number of collisions per incoming baryon. An important difference between the collisions of large, equal-sized nuclei and collisions of a very small nucleus with a large nucleus is that in the latter case the nucleons of the small nucleus collide with nucleons from the large nucleus that in almost all cases have not participated in the reaction before. This is not true in the first case, which is characterized by multiple collisions of both projectile and target nucleons. This difference can lead to differences in the stopping behavior in the reaction and could lead to the different observations in the two collision systems.

\section{Conclusion}

We have presented measurements of \Lam\ hyperon production in $d$ + Au collisions at $\sqrt{s_{NN}}$ = 200 GeV  at forward ($y = 2.75$) and backward ($y = -2.75$) rapidity. The comparison of minimum bias yields of \ALam\ and net \Lam\ to a variety of model calculations shows that the deuteron side is well described by all models used. On the gold side, only AMPT and EPOS are able to explain the net \Lam\ results, suggesting nuclear effects have an influence on hyperon production on the gold side of the collision. The observed centrality independence of the $\ALam /\Lam$ ratio shows that baryon number transport through associated production of strangeness and \ALam -\Lam\ pair production both have a similar dependence on the number of participants and the number of binary collisions in the reaction. The good agreement of the multichain model with the measured net \Lam\ yields using the same parameter as for lower energy data suggests energy independence of the nuclear stopping power for $p$($d$) + $A$ collisions over a wide range in center of mass energy.

\begin{acknowledgments} 
We thank L.W.~Chen, C.M.~Ko, V.~Topor-Pop and K.~Werner
for providing us the results for the different model calculations 
and M.~Gyulassy for help regarding the Multi-Chain Model.
We thank the RHIC Operations Group and RCF at BNL, and the
NERSC Center at LBNL for their support. This work was supported
in part by the Offices of NP and HEP within the U.S. DOE Office 
of Science; the U.S. NSF; the BMBF of Germany; CNRS/IN2P3, RA, RPL, and
EMN of France; EPSRC of the United Kingdom; FAPESP of Brazil;
the Russian Ministry of Science and Technology; the Ministry of
Education and the NNSFC of China; IRP and GA of the Czech Republic,
FOM of the Netherlands, DAE, DST, and CSIR of the Government
of India; Swiss NSF; the Polish State Committee for Scientific 
Research; SRDA of Slovakia, and the Korea Sci. \& Eng. Foundation.
\end{acknowledgments}

\bibliography{LambdaStopping}

\end{document}